# Mixed Chiral and Achiral Character in Substituted Ethane: A Next Generation QTAIM Perspective

Zi Li[1], Tianlv Xu[1], Herbert Früchtl[2], Tanja van Mourik[2], Steven R. Kirk[*1] and Samantha Jenkins[*1]

[1]*Key Laboratory of Chemical Biology and Traditional Chinese Medicine Research and Key Laboratory of Resource National and Local Joint Engineering Laboratory for New Petro-chemical Materials and Fine Utilization of Resources, College of Chemistry and Chemical Engineering, Hunan Normal University, Changsha, Hunan 410081, China*
[2]*EaStCHEM School of Chemistry, University of St Andrews, North Haugh, St Andrews, Fife KY16 9ST, Scotland, United Kingdom.*

email: steven.kirk@cantab.net
email: samanthajsuman@gmail.com

We use the newly introduced spanning stress tensor trajectory $\mathbb{U}_\sigma$-space construction within next generation quantum theory of atoms in molecules (NG-QTAIM) for a chirality investigation of singly and doubly substituted ethane with halogen substituents: F, Cl, Br. A lack of achiral character in $\mathbb{U}_\sigma$-space was discovered for singly substituted ethane. The resultant axial bond critical point (*BCP*) sliding responded more strongly to the increase in atomic number of the substituted halogen than the chirality. The presence of the very light F atom was found responsible for a very high degree of achiral character of the doubly substituted ethane.

# 1. Introduction

As early as 1848 Louis Pasteur proposed biomolecular homochirality as a possible simple 'chemical signature of life'[1]. More recently, Prelog[2] and Quack[3,4] considered the origins of homochirality. Quack concluded from extensive theoretical investigations that no direct relationship between energy difference and the left or right handedness of a chiral molecule existed, as may be expected for the L-amino acid dominant world. A chiral molecule generally has at least one chiral center and the Cahn–Ingold–Prelog (CIP) priority rules, which have recently been updated[5], are used to determine the chiral configuration (R/S) without calculations[6,7]. There are however situations where chirality cannot be assigned from the CIP rules, such as achiral molecules including optical isomers of coordination compounds used as stereoselectivity catalysts. The existence of chirality has important implications[2,8,9] for the origin of chiral asymmetry[10], which is one of the great mysteries in the understanding of the origin of life[11–18]. Distinguishing and understanding the chirality is essential to understanding the important role homochirality plays in life on earth[12,19,20].

A pair of S and R stereoisomers possess very similar physical properties but demonstrate strong enantiomeric preferences during chemical and biological reactions. Quantifying enantiomeric excess and handedness of chiral molecules therefore plays a crucial role in biochemistry and pharmacy[21–23]. As a result, development of ever more reliable methods to differentiate S and R stereoisomers is of current interest[24]. The recent work of Tremblay focuses on understanding the conditions required to modify the chirality during ultrafast electronic motion by bringing enantiomers out of equilibrium [25]. A selection of ultrashort linearly-polarised laser pulses were used to drive an ultrafast charge migration process by the excitation of a small number of low-lying excited states from the ground electronic state of $S_a$ and $R_a$ epoxypropane stereoisomers.

Recently, some of the current authors quantified a chirality-helicity measure[26] using next generation QTAIM (NG-QTAIM) [27]. This chirality-helicity measure is an association between molecular chirality and helical characteristics known as the chirality-helicity equivalence, introduced by Wang[28] that is consistent with photoexcitation circular dichroism experiments. Wang stated that the origin of this helical character was not provided solely by molecular geometries or attributable to steric effects, but would require future insight provided from the electronic structure. The interdependence of steric-electronic factors was recently discovered to be more complex[29] than was discernable from the molecular structures for the helical electronic transitions of spiroconjugated molecules[30,31].

Banerjee-Ghosh *et al.* demonstrated that charge density redistribution in chiral molecules, and not spatial effects[32], was responsible for an enantiospecific preference in electron spin orientation, consistent with Wang.

A recent investigation to quantify the chirality-helicity equivalence proposed by Wang discovered that what is generally understood as chirality is only *part* of the complete understanding of the behavior associated with 'chirality'. To address this realization we demonstrated that the bond-axiality, which is concerned with the motion of the bond critical point (*BCP*) along the torsion bond, should also be considered[31], in addition to the asymmetry of the chiral carbon, which is generally understood to be associated with chirality.

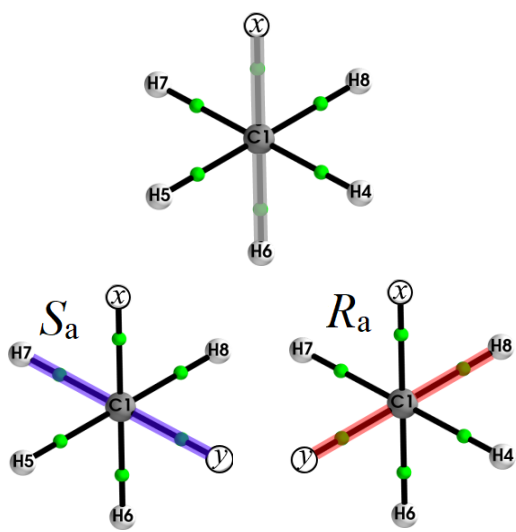

**Scheme 1**. Axial views down the torsion C1-C2 *BCP* bond-path of the molecular graph of singly substituted ethane (top) with *x* = F, Cl, Br, where *x* is bonded to the C1 atom. For the doubly substituted ethane the $S_a$-stereoisomer (bottom-left) and $R_a$-stereoisomer (bottom-right), the heaviest halogen atom is located at *x* and the lighter one at *y*; therefore the three pairs of $S_a$-stereoisomer and $R_a$-stereoisomer correspond to {*x* = Cl, *y* = F}, {*x* = Br, *y* = Cl}, {*x* = Br, *y* = F}. Note that *y* is replacing the H4 and H5 atom for the $S_a$-stereoisomer and $R_a$-stereoisomer, respectively, see **Table 1** and **Table 2**. The *x*, *y*, H4 and H5 atoms are bonded to the C1 atom and the H6, H7 and H8 atoms are bonded to the C2 atom.

Recently, we undertook an NG-QTAIM investigation into the chiral properties of ethane, where a new category of $\mathbb{U}_\sigma$-space chirality assignment was discovered with 'null' chirality, which we referred to as a **$Q_\sigma$** isomer in $\mathbb{U}_\sigma$-space [33]. We also discovered **$S_\sigma$** and **$R_\sigma$** stereoisomers in $\mathbb{U}_\sigma$-space and used this to explain the staggered conformation of ethane in terms of a preference for steric effects over hyper-conjugation.

In this investigation we will consider the NG-QTAIM perspective of the achiral and chiral behaviors of singly and doubly substituted ethane, see **Scheme 1**. We will consider the effect of increasing the atomic number of the halogen substituents, F, Cl and Br on the balance of the **$Q_\sigma$, $S_\sigma$** or **$R_\sigma$** $\mathbb{U}_\sigma$-space chirality assignments of the singly substituted ethane isomers as well as the doubly substituted ethane $S_a$- and $R_a$-geometric stereoisomers. Note we use the subscript "ₐ" to denote geometric stereoisomers to differentiate from the $\mathbb{U}_\sigma$-space stereoisomers **$S_\sigma$** and **$R_\sigma$**.

## 2. Theoretical Background and Computational Details

The essential background of QTAIM and next generation QTAIM (NG-QTAIM)[34–40] is provided in the **Supplementary Materials S1**, including the procedure to generate the stress tensor trajectories $\mathbb{T}_\sigma(s)$.

Bader's formulation of the stress tensor[41] is a standard option in the AIMAll QTAIM package [42] and is used in this investigation because of the superior performance of the stress tensor compared with the Hessian of $\rho(\mathbf{r})$ for distinguishing the $S_a$- and $R_a$-stereoisomers of lactic acid[43].

The chirality $\mathbb{C}_\sigma$ quantifies the bond torsion direction counter-clockwise (CCW) vs. clockwise (CW), i.e. the *circular* C1-C2 *BCP* displacement, where the largest magnitude stress tensor eigenvalue ($\lambda_{1\sigma}$) is associated with $\mathbf{e}_{1\sigma}$. The $\mathbf{e}_{1\sigma}$ corresponds to the direction in which the electrons at the C1-C2 *BCP* are subject to the most compressive forces, and therefore will be the direction along which the C1-C2 *BCP* electrons will be most displaced when the C1-C2 *BCP* is subjected to torsion[44]. The chirality $\mathbb{C}_\sigma$ is defined as the difference in the maximum projections (the dot product of the stress tensor $\mathbf{e}_{1\sigma}$ eigenvector and the *BCP* displacement $\mathbf{dr}$) of the $\mathbb{T}_\sigma(s)$ values between the CCW and CW torsion θ: $\mathbb{C}_\sigma = [(\mathbf{e}_{1\sigma}\cdot\mathbf{dr})_{max}]_{CCW} - [(\mathbf{e}_{1\sigma}\cdot\mathbf{dr})_{max}]_{CW}$. The bond-flexing $\mathbb{F}_\sigma$, defined as $\mathbb{F}_\sigma = [(\mathbf{e}_{2\sigma}\cdot\mathbf{dr})_{max}]_{CCW} - [(\mathbf{e}_{2\sigma}\cdot\mathbf{dr})_{max}]_{CW}$, provides a measure of the 'flexing-strain' of a bond-path. The bond-axiality $\mathbb{A}_\sigma$, which provides a measure of the chiral asymmetry, is defined as $\mathbb{A}_\sigma = [(\mathbf{e}_{3\sigma}\cdot\mathbf{dr})_{max}]_{CCW} - [(\mathbf{e}_{3\sigma}\cdot\mathbf{dr})_{max}]_{CW}$. The bond-axiality $\mathbb{A}_\sigma$ quantifies the direction of *axial* displacement of the bond critical point (*BCP*) in response to the bond torsion (CCW or CW), i.e. the sliding of the *BCP* along the bond-path[45]. The signs of the chirality $\mathbb{C}_\sigma$, bond-flexing $\mathbb{F}_\sigma$ and bond-axiality $\mathbb{A}_\sigma$ determine the prevalence of $\mathbf{S}_\sigma$ ($\mathbb{C}_\sigma > 0$, $\mathbb{F}_\sigma > 0$, $\mathbb{A}_\sigma > 0$) or $\mathbf{R}_\sigma$ ($\mathbb{C}_\sigma < 0$, $\mathbb{F}_\sigma < 0$, $\mathbb{A}_\sigma < 0$) character, see **Table 1** and **Table 2**. The intermediate results for the $\mathbb{U}_\sigma$-space distortion sets {$\mathbb{C}_\sigma,\mathbb{F}_\sigma,\mathbb{A}_\sigma$} are provided in the **Supplementary Materials S5.**

An additional null-chirality assignment $\mathbf{Q}_\sigma$ (≈ 0 chiral character) occurs where a plot of ellipticity ε vs. torsion θ displays CCW and CW portions that are symmetrical about torsion θ = 0.0º. The ± sign is therefore not used with the chirality $\mathbb{C}_\sigma$ assignment $\mathbf{Q}_\sigma$ as it is for the $\mathbf{S}_\sigma$ and $\mathbf{R}_\sigma$ assignments, where for the latter mirror symmetry is only present for the $\mathbf{S}_\sigma$ CCW vs. $\mathbf{R}_\sigma$ CW and $\mathbf{S}_\sigma$ CW vs. $\mathbf{R}_\sigma$ CCW plots of ellipticity ε vs. torsion θ.

We include all contributions to the $\mathbb{U}_\sigma$-space chirality from the 'chiral' centre C1 by considering the entire bonding environment of the 'chiral' carbon atom (C1) by constructing all nine torsion C1-C2 *BCP* $\mathbb{T}_\sigma(s)$ using dihedral angles that include the C1 atom, see **Scheme 1** and the Computational Details section. We refer to this process of using all nine torsion C1-C2 *BCP* $\mathbb{T}_\sigma(s)$ as the spanning $\mathbb{U}_\sigma$-space chirality construction. The NG-QTAIM interpretation of the singly and doubly substituted ethane as either an achiral or chiral molecule requires calculation of all the symmetry inequivalent stress tensor trajectories $\mathbb{T}_\sigma(s)$ through the torsion C1-C2 *BCP*. This results in a complete set of ethane $\mathbb{U}_\sigma$-space isomers, with possible chirality assignments $\mathbf{Q}_\sigma$, $\mathbf{S}_\sigma$ or $\mathbf{R}_\sigma$.

The linear sum of the individual components of the symmetry inequivalent $\mathbb{U}_\sigma$-space distortion sets $\sum${$\mathbb{C}_\sigma$, $\mathbb{F}_\sigma$, $\mathbb{A}_\sigma$} will be calculated to provide the resultant chiral character for the singly and doubly substituted ethane molecular graph. Then $\mathbb{C}_{helicity}$ (= $\mathbb{C}_\sigma|\mathbb{A}_\sigma|$) of the resultant linear sum $\sum${$\mathbb{C}_\sigma$, $\mathbb{F}_\sigma$, $\mathbb{A}_\sigma$} in addition to the sum of

the individual $\mathbb{C}_{helicity}$; values of $\sum\mathbb{C}_{helicity} \approx 0$ correspond to the null-chirality assignment $\mathbf{Q_\sigma}$, see **Table 1** and **Table 2**. In this investigation the presence of a mix of $\mathbf{S_\sigma}$ and $\mathbf{R_\sigma}$ chirality assignments for an $S_a$ or $R_a$ stereoisomer will be referred to as *mixed* chirality in $\mathbb{U}_\sigma$-space. An equal mixing of the $\mathbf{S_\sigma}$ and $\mathbf{R_\sigma}$ chirality for the $S_a$ and $R_a$ geometric stereoisomers, as determined by the value of $\sum_{\mathbf{S\sigma}}\{\mathbb{C}_\sigma\}/|\sum_{\mathbf{R\sigma}}\{\mathbb{C}_\sigma\}|\} = 1$, would correspond to pure achiral character present in $\mathbb{U}_\sigma$-space for each of the $S_a$ and $R_a$ geometric stereoisomers.

*Computational Details*

The singly and doubly substituted ethane molecular structures were initially geometry-optimized with 'verytight' convergence criteria at the B3LYP/cc-pVQZ level of DFT theory using Gaussian 09.E01 [46] with an 'ultrafine' integration grid. The wavefunctions were converged to < $10^{-10}$ RMS change in the density matrix and < $10^{-8}$ maximum change in the density matrix. We determine the direction of torsion as CCW (0.0º ≤ θ ≤ +180.0º) or CW (-180.0º ≤ θ ≤ 0.0º) from an increase or a decrease in the dihedral angle, respectively. The $\mathbb{T}_\sigma(s)$ for all nine possible ordered sets of four atoms that define the dihedral angle {(3126, 3127, 3128), (4126, 4127, 4128), (5126, 5127, 5128)} are calculated. The dihedral atom numbering is provided in **Scheme 1**.

Single-point calculations were then undertaken on each torsion scan geometry, where the SCF iterations were converged to < $10^{-10}$ RMS change in the density matrix and < $10^{-8}$ maximum change in the density matrix to yield the final wavefunctions for analysis. QTAIM and stress tensor analysis was performed with the AIMAll[42] and QuantVec[47] suite on each wavefunction obtained in the previous step. All molecular graphs were additionally confirmed to be free of non-nuclear attractor critical points.

## 3. Results and discussions

The scalar measures of distance, relative energy ΔE and ellipticity ε used in this investigation for the singly and doubly substituted ethane are insufficient to explain any chiral effects and are provided in the **Supplementary Materials S2-S4** respectively.

Previously, an investigation into ethane [33] subjected to an applied electric field used a representative selection of three $\mathbb{U}_\sigma$-space distortion sets $\{\mathbb{C}_\sigma, \mathbb{F}_\sigma, \mathbb{A}_\sigma\}$ of the C1-C2 *BCP* $\mathbb{T}_\sigma(s)$. In this investigation on singly and doubly substituted ethane we will use all nine possible $\mathbb{U}_\sigma$-space distortion sets $\{\mathbb{C}_\sigma, \mathbb{F}_\sigma, \mathbb{A}_\sigma\}$, see **Scheme 1**, **Table 1** and **Table 2**.

The stress tensor trajectory $\mathbb{T}_\sigma(s)$ plots for the singly (F, Cl, Br) substituted ethane are rather similar to each other and therefore are provided in the **Supplementary Materials S5**. The NG-QTAIM interpretation of the chirality of the singly substituted (F or Cl or Br) ethane is that ethane is an achiral molecule with constituent chirality assignments $\mathbf{Q_\sigma}$, $\mathbf{S_\sigma}$ or $\mathbf{R_\sigma}$, see **Scheme 1** and **Table 1**. This result was obtained by including all nine possible torsion C1-C2 *BCP* stress tensor trajectories $\mathbb{T}_\sigma(s)$. Note, for comparison, for the pure ethane molecule the chirality helicity function $\mathbb{C}_{helicity} \approx 0$, +0.0003 and -0.0003 for the $\mathbf{Q}_\sigma$, $\mathbf{S}_\sigma$ and $\mathbf{R}_\sigma$ stereoisomers in $\mathbb{U}_\sigma$-space, respectively [33].

The $\mathbb{T}_\sigma(s)$ plots for the doubly substituted ethane are provided in **Figure 1**. For the singly substituted ethane the isomer pairs: ($D_{3127}$ and $D_{3128}$), ($D_{4128}$ and $D_{5127}$) and ($D_{4127}$ and $D_{5128}$) each possess opposite $\mathbf{S}_\sigma$ and $\mathbf{R}_\sigma$ chirality assignments with approximately equal magnitudes of the corresponding elements of the distortion sets {$\mathbb{C}_\sigma$, $\mathbb{F}_\sigma$, $\mathbb{A}_\sigma$}. Therefore, the ($D_{3127}$ and $D_{3128}$), ($D_{4128}$ and $D_{5127}$) and ($D_{4127}$ and $D_{5128}$) pairs of isomers can be regarded as stereoisomers in $\mathbb{U}_\sigma$-space. For the singly substituted ethane the $D_{3126}$, $D_{4126}$ and $D_{5126}$ isomers are each indicated as possessing chirality assignment $\mathbf{Q}_\sigma$. This assignment ($\mathbf{Q}_\sigma$) corresponding to null-chirality on the basis of the very low values of the chirality $\mathbb{C}_\sigma$ and bond-axiality $\mathbb{B}_\sigma$, i.e. an order of magnitude lower than for the $\mathbf{S}_\sigma$ or $\mathbf{R}_\sigma$. The linear sum of the distortion sets $\sum\{\mathbb{C}_\sigma, \mathbb{F}_\sigma, \mathbb{A}_\sigma\} = 0$ for the complete set of nine isomers {(3126, 3127, 3128), (4126, 4127, 4128), (5126, 5127, 5128)} corresponds to a chirality assignment $\mathbf{Q}_\sigma$ that we refer to as 'null-chirality', consistent with the singly substituted ethane being formally achiral. The sum of the chirality-helicity function $\sum\mathbb{C}_{helicity} \approx 0$ for the complete set of nine distortion sets {$\mathbb{C}_\sigma$, $\mathbb{F}_\sigma$, $\mathbb{A}_\sigma$} also corresponds to a chirality assignment $\mathbf{Q}_\sigma$. The sums of the $\mathbf{S}_\sigma$ components $\sum_{\mathbf{S}\sigma}\{\mathbb{C}_\sigma, \mathbb{A}_\sigma\}$ and $\mathbf{R}_\sigma$ components $\sum_{\mathbf{R}\sigma}\{\mathbb{C}_\sigma, \mathbb{A}_\sigma\}$ are approximately equal in magnitude. This demonstrates the stereoisomer character of the singly substituted ethane in $\mathbb{U}_\sigma$-space. Additionally, the $\sum_{\mathbf{S}\sigma}\{\mathbb{C}_\sigma, \mathbb{A}_\sigma\}$ increase with atomic number (F, Cl, Br) with the strongest correlation for $\mathbb{A}_\sigma$.

Examination of the doubly substituted ethane for the $S_a$ and $R_a$ geometric stereoisomers demonstrates agreement with the CIP rules for the largest value of $\sum_{\mathbf{S}\sigma}\{\mathbb{C}_\sigma\}$ or $\sum_{\mathbf{R}\sigma}\{\mathbb{C}_\sigma\}$, respectively, of the $S_a$ and $R_a$ stereoisomers. In particular, the separate contributions of the chirality $\mathbb{C}_\sigma$ of the $S_a$ geometric stereoisomer $\sum_{\mathbf{S}\sigma}\{\mathbb{C}_\sigma\} = 1.898$ and $\sum_{\mathbf{R}\sigma}\{C_\sigma\} = -0.9977$, see **Table 2**. The sum over all nine isomers $\sum_{\mathbf{S}\sigma,\mathbf{R}\sigma}\{\mathbb{C}_\sigma, \mathbb{F}_\sigma, \mathbb{A}_\sigma\}$ results in equal magnitudes of each of the $\mathbb{C}_\sigma$, $\mathbb{F}_\sigma$ and $\mathbb{A}_\sigma$, e.g. for the F-Cl ethane: $\mathbb{C}_\sigma = 0.901$ [$\mathbf{S}_\sigma$] and -0.901[$\mathbf{R}_\sigma$] for the $S_a$ and $R_a$ geometric stereoisomers respectively, thus demonstrating stereoisomer characteristics in $\mathbb{U}_\sigma$-space. The sums of the chirality helicity function $\mathbb{C}_{helicity}$, $\sum\mathbb{C}_{helicity}$, possess equal magnitudes, but opposite sign, for the $S_a$ and $R_a$ geometric stereoisomers. In other words, the $S_a$ geometric stereoisomer is demonstrated to comprise a *mix* of chirality contributions, dominated by contributions from the $\mathbf{S}_\sigma$, but containing a sizable contribution with a $\mathbf{R}_\sigma$ chirality assignment. The converse is true for the $R_a$ geometric stereoisomer. The presence of a mix of $\mathbf{S}_\sigma$ and $\mathbf{R}_\sigma$ chirality for each of the $S_a$ and $R_a$ geometric stereoisomers is referred to as mixed chirality in $\mathbb{U}_\sigma$-space. The doubly substituted ethane with the lowest mixed chirality in $\mathbb{U}_\sigma$-space is Cl-Br-ethane, where $\sum_{\mathbf{S}\sigma}\{\mathbb{C}_\sigma\}/|\sum_{\mathbf{R}\sigma}\{\mathbb{C}_\sigma\}|\} \approx 12$. In contrast, F-Br-ethane and Br-F-ethane both possess $\sum_{\mathbf{S}\sigma}\{\mathbb{C}_\sigma\}/|\sum_{\mathbf{R}\sigma}\{\mathbb{C}_\sigma\}|\} \approx 2$, see **Table 2**.

**Table 1**. The singly substituted ethane torsion C1-C2 *BCP* $\mathbb{U}_\sigma$-space distortion sets {chirality $\mathbb{C}_\sigma$, bond-flexing $\mathbb{F}_\sigma$, bond-axiality $\mathbb{A}_\sigma$}, sum of the nine isomers: $\sum_{S\sigma,R\sigma}\{\mathbb{C}_\sigma,\mathbb{F}_\sigma,\mathbb{A}_\sigma\}$, sum of the $\mathbf{S}_\sigma$ components $\sum_{S\sigma}\{\mathbb{C}_\sigma,\mathbb{A}_\sigma\}$ and $\mathbf{R}_\sigma$ components $\sum_{R\sigma}\{\mathbb{C}_\sigma,\mathbb{A}_\sigma\}$, chirality-helicity function $\mathbb{C}_{helicity} = (\mathbb{C}_\sigma)(|\mathbb{A}_\sigma|)$ and the sum of $\mathbb{C}_{helicity}$ over the nine isomers $\sum\mathbb{C}_{helicity}$. The four-digit sequence (left column) refers to the atom numbering used in the dihedral angles to construct the stress tensor trajectories $\mathbb{T}_\sigma(s)$ correspond to the $D_{3126}$, $D_{4128}$, $D_{5127}$, ... etc. isomer names, see **Scheme 1**.

| Isomer | $\{\mathbb{C}_\sigma, \mathbb{F}_\sigma, \mathbb{A}_\sigma\}$ | $\mathbb{C}_{helicity}$ | $[\mathbb{C}_\sigma,\mathbb{A}_\sigma]$ |
|---|---|---|---|
| *F-ethane* | | | |
| $D_{3126}$ | {-0.000014[$\mathbf{R}_\sigma$], -0.000002[$\mathbf{R}_\sigma$], -0.000022[$\mathbf{R}_\sigma$]} | $\approx 0$ (-3.14×10$^{-10}$) | $[\mathbf{Q}_\sigma]$ |
| $D_{3127}$ | {0.167219[$\mathbf{S}_\sigma$], -0.293389[$\mathbf{R}_\sigma$], 0.005542[$\mathbf{S}_\sigma$]} | 0.0009 | $[\mathbf{S}_\sigma,\mathbf{S}_\sigma]$ |
| $D_{3128}$ | {-0.167170[$\mathbf{R}_\sigma$], 0.292479[$\mathbf{S}_\sigma$], -0.005563[$\mathbf{R}_\sigma$]} | -0.0009 | $[\mathbf{R}_\sigma,\mathbf{R}_\sigma]$ |
| $D_{4126}$ | {0.007700[$\mathbf{S}_\sigma$], -0.019744[$\mathbf{R}_\sigma$], 0.003671[$\mathbf{S}_\sigma$]} | $\approx 0$ (2.83×10$^{-5}$) | $[\mathbf{Q}_\sigma]$ |
| $D_{4127}$ | {0.235443[$\mathbf{S}_\sigma$], -0.291961[$\mathbf{R}_\sigma$], -0.001050[$\mathbf{R}_\sigma$]} | 0.0002 | $[\mathbf{S}_\sigma,\mathbf{R}_\sigma]$ |
| $D_{4128}$ | {-0.203761[$\mathbf{R}_\sigma$], 0.237553[$\mathbf{S}_\sigma$], -0.008936[$\mathbf{R}_\sigma$]} | -0.0018 | $[\mathbf{R}_\sigma,\mathbf{R}_\sigma]$ |
| $D_{5126}$ | {-0.007692[$\mathbf{R}_\sigma$], 0.019733[$\mathbf{S}_\sigma$], -0.003683[$\mathbf{R}_\sigma$]} | $\approx 0$ (-2.83×10$^{-5}$) | $[\mathbf{Q}_\sigma]$ |
| $D_{5127}$ | {0.203734[$\mathbf{S}_\sigma$], -0.238685[$\mathbf{R}_\sigma$], 0.008958[$\mathbf{S}_\sigma$]} | 0.0018 | $[\mathbf{S}_\sigma,\mathbf{S}_\sigma]$ |
| $D_{5128}$ | {-0.235470[$\mathbf{R}_\sigma$], 0.291163[$\mathbf{S}_\sigma$], 0.001052[$\mathbf{S}_\sigma$]} | -0.0002 | $[\mathbf{R}_\sigma,\mathbf{S}_\sigma]$ |
| | $\sum_{S\sigma,R\sigma}\{\mathbb{C}_\sigma, \mathbb{F}_\sigma, \mathbb{A}_\sigma\}$ | $\sum\mathbb{C}_{helicity}$ | |
| | {-0.00001[$\mathbf{R}_\sigma$], -0.0029[$\mathbf{R}_\sigma$], -0.00003[$\mathbf{R}_\sigma$]} | 0 | $[\mathbf{Q}_\sigma]$ |
| | $\sum_{S\sigma}\{\mathbb{C}_\sigma,\mathbb{A}_\sigma\}$ | $\sum_{R\sigma}\{\mathbb{C}_\sigma,\mathbb{A}_\sigma\}$ | |
| | {0.6141[$\mathbf{S}_\sigma$], 0.0192[$\mathbf{S}_\sigma$]} | {-0.6141[$\mathbf{R}_\sigma$], -0.0193[$\mathbf{R}_\sigma$]} | |
| *Cl-ethane* | | | |
| $D_{3126}$ | {-0.000001[$\mathbf{R}_\sigma$], -0.000009[$\mathbf{R}_\sigma$], 0.000006[$\mathbf{S}_\sigma$]} | $\approx 0$ (-7.65×10$^{-12}$) | $[\mathbf{Q}_\sigma]$ |
| $D_{3127}$ | {0.530364[$\mathbf{S}_\sigma$], -0.153082[$\mathbf{R}_\sigma$], -0.015103[$\mathbf{R}_\sigma$]} | 0.0080 | $[\mathbf{S}_\sigma,\mathbf{R}_\sigma]$ |
| $D_{3128}$ | {-0.530402[$\mathbf{R}_\sigma$], 0.152974[$\mathbf{S}_\sigma$], 0.015084[$\mathbf{S}_\sigma$]} | -0.0080 | $[\mathbf{R}_\sigma,\mathbf{S}_\sigma]$ |
| $D_{4126}$ | {-0.058372[$\mathbf{R}_\sigma$], -0.112844[$\mathbf{R}_\sigma$], 0.001391[$\mathbf{S}_\sigma$]} | $\approx 0$ (-8.12×10$^{-5}$) | $[\mathbf{Q}_\sigma]$ |
| $D_{4127}$ | {0.525440[$\mathbf{S}_\sigma$], -0.156354[$\mathbf{R}_\sigma$], -0.019219[$\mathbf{R}_\sigma$]} | 0.0101 | $[\mathbf{S}_\sigma,\mathbf{R}_\sigma]$ |
| $D_{4128}$ | {-0.412889[$\mathbf{R}_\sigma$], 0.129050[$\mathbf{S}_\sigma$], 0.014041[$\mathbf{S}_\sigma$]} | -0.0058 | $[\mathbf{R}_\sigma,\mathbf{S}_\sigma]$ |
| $D_{5126}$ | {0.058388[$\mathbf{S}_\sigma$], 0.112880[$\mathbf{S}_\sigma$], -0.001392[$\mathbf{R}_\sigma$]} | $\approx 0$ (-8.13×10$^{-5}$) | $[\mathbf{Q}_\sigma]$ |
| $D_{5127}$ | {0.413996[$\mathbf{S}_\sigma$], -0.127786[$\mathbf{R}_\sigma$], -0.014039[$\mathbf{R}_\sigma$]} | 0.0058 | $[\mathbf{S}_\sigma,\mathbf{R}_\sigma]$ |
| $D_{5128}$ | {-0.525488[$\mathbf{R}_\sigma$], 0.156356[$\mathbf{S}_\sigma$], 0.019231[$\mathbf{S}_\sigma$]} | -0.0101 | $[\mathbf{R}_\sigma,\mathbf{S}_\sigma]$ |
| | $\sum_{S\sigma,R\sigma}\{\mathbb{C}_\sigma, \mathbb{F}_\sigma, \mathbb{A}_\sigma\}$ | $\sum\mathbb{C}_{helicity}$ | |
| | {0.0010[$\mathbf{S}_\sigma$], 0.0012[$\mathbf{S}_\sigma$], 0.000006[$\mathbf{S}_\sigma$]} | 0 | $[\mathbf{Q}_\sigma]$ |
| | $\sum_{S\sigma}\{\mathbb{C}_\sigma,\mathbb{A}_\sigma\}$ | $\sum_{R\sigma}\{\mathbb{C}_\sigma,\mathbb{A}_\sigma\}$ | |
| | {1.5282[$\mathbf{S}_\sigma$], 0.0498[$\mathbf{S}_\sigma$]} | {-1.5272[$\mathbf{R}_\sigma$], -0.0498[$\mathbf{R}_\sigma$]} | |
| *Br-ethane* | | | |
| $D_{3126}$ | {-0.000049[$\mathbf{R}_\sigma$], 0.000013[$\mathbf{S}_\sigma$], -0.000012[$\mathbf{R}_\sigma$]} | $\approx 0$ (-5.84×10$^{-10}$) | $[\mathbf{Q}_\sigma]$ |
| $D_{3127}$ | {0.583865[$\mathbf{S}_\sigma$], -0.159965[$\mathbf{R}_\sigma$], -0.025146[$\mathbf{R}_\sigma$]} | 0.0147 | $[\mathbf{S}_\sigma,\mathbf{R}_\sigma]$ |
| $D_{3128}$ | {-0.583985[$\mathbf{R}_\sigma$], 0.160061[$\mathbf{S}_\sigma$], 0.025172[$\mathbf{S}_\sigma$]} | -0.0147 | $[\mathbf{R}_\sigma,\mathbf{S}_\sigma]$ |
| $D_{4126}$ | {-0.079245[$\mathbf{R}_\sigma$], -0.154251[$\mathbf{R}_\sigma$], 0.001230[$\mathbf{S}_\sigma$]} | $\approx 0$ (-9.74×10$^{-5}$) | $[\mathbf{Q}_\sigma]$ |
| $D_{4127}$ | {0.540902[$\mathbf{S}_\sigma$], -0.166512[$\mathbf{R}_\sigma$], -0.028603[$\mathbf{R}_\sigma$]} | 0.0155 | $[\mathbf{S}_\sigma,\mathbf{R}_\sigma]$ |
| $D_{4128}$ | {-0.387654[$\mathbf{R}_\sigma$], 0.162074[$\mathbf{S}_\sigma$], 0.027027[$\mathbf{S}_\sigma$]} | -0.0105 | $[\mathbf{R}_\sigma,\mathbf{S}_\sigma]$ |
| $D_{5126}$ | {0.079315[$\mathbf{S}_\sigma$], 0.154404[$\mathbf{S}_\sigma$], -0.001227[$\mathbf{R}_\sigma$]} | $\approx 0$ (-9.73×10$^{-5}$) | $[\mathbf{Q}_\sigma]$ |
| $D_{5127}$ | {0.387677[$\mathbf{S}_\sigma$], -0.163163[$\mathbf{R}_\sigma$], -0.027016[$\mathbf{R}_\sigma$]} | 0.0105 | $[\mathbf{S}_\sigma,\mathbf{R}_\sigma]$ |
| $D_{5128}$ | {-0.540781[$\mathbf{R}_\sigma$], 0.166429[$\mathbf{S}_\sigma$], 0.028581[$\mathbf{S}_\sigma$]} | -0.0155 | $[\mathbf{R}_\sigma,\mathbf{S}_\sigma]$ |
| | $\sum_{S\sigma,R\sigma}\{\mathbb{C}_\sigma, \mathbb{F}_\sigma, \mathbb{A}_\sigma\}$ | $\sum\mathbb{C}_{helicity}$ | |
| | {0.00005[$\mathbf{S}_\sigma$], -0.0009[$\mathbf{R}_\sigma$], 0.000006[$\mathbf{S}_\sigma$]} | 0 | $[\mathbf{Q}_\sigma]$ |
| | $\sum_{S\sigma}\{\mathbb{C}_\sigma,\mathbb{A}_\sigma\}$ | $\sum_{R\sigma}\{\mathbb{C}_\sigma,\mathbb{A}_\sigma\}$ | |
| | {1.5918[$\mathbf{S}_\sigma$], 0.0820[$\mathbf{S}_\sigma$]} | {-1.5917[$\mathbf{R}_\sigma$], -0.0820[$\mathbf{R}_\sigma$]} | |

**Table 2**. The doubly substituted ethane torsion C1-C2 *BCP* $\mathbb{U}_\sigma$-space distortion sets {$\mathbb{C}_\sigma$, $\mathbb{F}_\sigma$, $\mathbb{A}_\sigma$} and the $\mathbb{U}_\sigma$-space achirality ratio $\sum_{\mathbf{S\sigma}}\{\mathbb{C}_\sigma\}/|\sum_{\mathbf{R\sigma}}\{\mathbb{C}_\sigma\}|$. Refer to the heading of **Table 1** for further details. Note that $S_a$ and $R_a$ correspond to the geometric stereoisomers, see **Scheme 1**.

| | $S_a$ | | | | $R_a$ | | | |
|---|---|---|---|---|---|---|---|---|
| Isomer | {$\mathbb{C}_\sigma$, $\mathbb{F}_\sigma$, $\mathbb{A}_\sigma$} | | $\mathbb{C}_{helicity}$ | [$\mathbb{C}_\sigma$,$\mathbb{A}_\sigma$] | {$\mathbb{C}_\sigma$, $\mathbb{F}_\sigma$, $\mathbb{A}_\sigma$} | | $\mathbb{C}_{helicity}$ | [$\mathbb{C}_\sigma$,$\mathbb{A}_\sigma$] |

*F-Cl-ethane*

| | | | | | | | | |
|---|---|---|---|---|---|---|---|---|
| $D_{3126}$ | {0.142058[**S**$_\sigma$], 0.098417[**S**$_\sigma$], 0.023993[**S**$_\sigma$]} | | 0.0034 | [**S**$_\sigma$,**S**$_\sigma$] | {-0.142118[**R**$_\sigma$], -0.099095[**R**$_\sigma$], -0.023983[**R**$_\sigma$]} | | -0.0034 | [**R**$_\sigma$,**R**$_\sigma$] |
| $D_{3127}$ | {0.544855[**S**$_\sigma$], 0.070844[**S**$_\sigma$], -0.032938[**R**$_\sigma$]} | | 0.0179 | [**S**$_\sigma$,**R**$_\sigma$] | {0.324119[**S**$_\sigma$], 0.583893[**S**$_\sigma$], -0.030021[**R**$_\sigma$]} | | 0.0097 | [**S**$_\sigma$,**R**$_\sigma$] |
| $D_{3128}$ | {-0.323278[**R**$_\sigma$], -0.584156[**R**$_\sigma$], 0.029974[**S**$_\sigma$]} | | -0.0097 | [**R**$_\sigma$,**S**$_\sigma$] | {-0.545013[**R**$_\sigma$], -0.070787[**R**$_\sigma$], 0.032975[**S**$_\sigma$]} | | -0.0180 | [**R**$_\sigma$,**S**$_\sigma$] |
| $D_{4126}$ | {0.140060[**S**$_\sigma$], 0.109472[**S**$_\sigma$], 0.027200[**S**$_\sigma$]} | | 0.0038 | [**S**$_\sigma$,**S**$_\sigma$] | {-0.119493[**R**$_\sigma$], -0.224093[**R**$_\sigma$], -0.028328[**R**$_\sigma$]} | | -0.0034 | [**R**$_\sigma$,**R**$_\sigma$] |
| $D_{4127}$ | {0.532049[**S**$_\sigma$], 0.095174[**S**$_\sigma$], -0.032489[**R**$_\sigma$]} | | 0.0173 | [**S**$_\sigma$,**R**$_\sigma$] | {0.362047[**S**$_\sigma$], 0.574724[**S**$_\sigma$], -0.023964[**R**$_\sigma$]} | | 0.0087 | [**S**$_\sigma$,**R**$_\sigma$] |
| $D_{4128}$ | {-0.312172[**R**$_\sigma$], -0.600659[**R**$_\sigma$], 0.026774[**S**$_\sigma$]} | | -0.0084 | [**R**$_\sigma$,**S**$_\sigma$] | {-0.419983[**R**$_\sigma$], 0.028998[**S**$_\sigma$], 0.035868[**S**$_\sigma$]} | | -0.0151 | [**R**$_\sigma$,**S**$_\sigma$] |
| $D_{5126}$ | {0.119437[**S**$_\sigma$], 0.223768[**S**$_\sigma$], 0.027831[**S**$_\sigma$]} | | 0.0033 | [**S**$_\sigma$,**S**$_\sigma$] | {-0.140219[**R**$_\sigma$], -0.109432[**R**$_\sigma$], -0.027182[**R**$_\sigma$]} | | -0.0038 | [**R**$_\sigma$,**R**$_\sigma$] |
| $D_{5127}$ | {0.419988[**S**$_\sigma$], -0.029150[**R**$_\sigma$], -0.035837[**R**$_\sigma$]} | | 0.0151 | [**S**$_\sigma$,**R**$_\sigma$] | {0.312187[**S**$_\sigma$], 0.600659[**S**$_\sigma$], -0.026766[**R**$_\sigma$]} | | 0.0084 | [**S**$_\sigma$,**R**$_\sigma$] |
| $D_{5128}$ | {-0.362224[**R**$_\sigma$], -0.574683[**R**$_\sigma$], 0.023966[**S**$_\sigma$]} | | -0.0087 | [**R**$_\sigma$,**S**$_\sigma$] | {-0.532089[**R**$_\sigma$], -0.095057[**R**$_\sigma$], 0.032494[**S**$_\sigma$]} | | -0.0173 | [**R**$_\sigma$,**S**$_\sigma$] |

$\sum_{\mathbf{S\sigma,R\sigma}}\{\mathbb{C}_\sigma, \mathbb{F}_\sigma, \mathbb{A}_\sigma\}$  $\sum\mathbb{C}_{helicity}$  $\sum_{\mathbf{S\sigma,R\sigma}}\{\mathbb{C}_\sigma, \mathbb{F}_\sigma, \mathbb{A}_\sigma\}$  $\sum\mathbb{C}_{helicity}$

{0.9008 [**S**$_\sigma$], -1.1910 [**R**$_\sigma$], 0.0585 [**S**$_\sigma$]}  0.0340  [**S**$_\sigma$,**S**$_\sigma$]  {-0.9006 [**R**$_\sigma$], 1.1898 [**S**$_\sigma$], -0.0589 [**R**$_\sigma$]}  -0.0342  [**R**$_\sigma$,**R**$_\sigma$]

$\sum_{\mathbf{S\sigma}}\{\mathbb{C}_\sigma,\mathbb{A}_\sigma\}$  $\sum_{\mathbf{R\sigma}}\{\mathbb{C}_\sigma,\mathbb{A}_\sigma\}$  $\sum_{\mathbf{S\sigma}}\{\mathbb{C}_\sigma,\mathbb{A}_\sigma\}$  $\sum_{\mathbf{R\sigma}}\{\mathbb{C}_\sigma,\mathbb{A}_\sigma\}$

{1.8984 [**S**$_\sigma$], 0.1597 [**S**$_\sigma$]}  {-0.9977 [**R**$_\sigma$], -0.1013 [**R**$_\sigma$]}  {0.9984 [**S**$_\sigma$], 0.1013 [**S**$_\sigma$]}  {-1.8989 [**R**$_\sigma$], -0.1602 [**R**$_\sigma$]}

$\sum_{\mathbf{S\sigma}}\{\mathbb{C}_\sigma\}/|\sum_{\mathbf{R\sigma}}\{\mathbb{C}_\sigma\}|\}$ |$S_a$ = 1.9028

*Cl-Br-ethane*

| | | | | | | | | |
|---|---|---|---|---|---|---|---|---|
| $D_{3126}$ | {0.142347[**S**$_\sigma$], -0.126939[**R**$_\sigma$], 0.030052[**S**$_\sigma$]} | | 0.0043 | [**S**$_\sigma$,**S**$_\sigma$] | {-0.142405[**R**$_\sigma$], 0.125579[**S**$_\sigma$], -0.030034[**R**$_\sigma$]} | | -0.0043 | [**R**$_\sigma$,**R**$_\sigma$] |
| $D_{3127}$ | {0.810931[**S**$_\sigma$], 0.337552[**S**$_\sigma$], -0.053673[**R**$_\sigma$]} | | 0.0435 | [**S**$_\sigma$,**R**$_\sigma$] | {0.083099[**S**$_\sigma$], 0.341168[**S**$_\sigma$], -0.004421[**R**$_\sigma$]} | | 0.0004 | [**S**$_\sigma$,**R**$_\sigma$] |
| $D_{3128}$ | {-0.083639[**R**$_\sigma$], -0.341316[**R**$_\sigma$], 0.004413[**S**$_\sigma$]} | | -0.0004 | [**R**$_\sigma$,**S**$_\sigma$] | {-0.810976[**R**$_\sigma$], -0.338399[**R**$_\sigma$], 0.053740[**S**$_\sigma$]} | | -0.0436 | [**R**$_\sigma$,**S**$_\sigma$] |
| $D_{4126}$ | {0.085004[**S**$_\sigma$], -0.042120[**R**$_\sigma$], 0.034848[**S**$_\sigma$]} | | 0.0030 | [**S**$_\sigma$,**S**$_\sigma$] | {-0.061939[**R**$_\sigma$], -0.308720[**R**$_\sigma$], -0.042080[**R**$_\sigma$]} | | -0.0026 | [**R**$_\sigma$,**R**$_\sigma$] |
| $D_{4127}$ | {0.804689[**S**$_\sigma$], 0.410931[**S**$_\sigma$], -0.050652[**R**$_\sigma$]} | | 0.0408 | [**S**$_\sigma$,**R**$_\sigma$] | {0.114210[**S**$_\sigma$], 0.315839[**S**$_\sigma$], -0.004005[**R**$_\sigma$]} | | 0.0005 | [**S**$_\sigma$,**R**$_\sigma$] |
| $D_{4128}$ | {0.037226[**S**$_\sigma$], -0.221889[**R**$_\sigma$], 0.003077[**S**$_\sigma$]} | | 0.0001 | [**S**$_\sigma$,**S**$_\sigma$] | {-0.471592[**R**$_\sigma$], 0.075689[**S**$_\sigma$], 0.064420[**S**$_\sigma$]} | | -0.0304 | [**R**$_\sigma$,**S**$_\sigma$] |
| $D_{5126}$ | {0.061856[**S**$_\sigma$], 0.308749[**S**$_\sigma$], 0.042072[**S**$_\sigma$]} | | 0.0026 | [**S**$_\sigma$,**S**$_\sigma$] | {-0.085311[**R**$_\sigma$], 0.041815[**S**$_\sigma$], -0.034836[**R**$_\sigma$]} | | -0.0030 | [**R**$_\sigma$,**R**$_\sigma$] |
| $D_{5127}$ | {0.471413[**S**$_\sigma$], -0.074717[**R**$_\sigma$], -0.064441[**R**$_\sigma$]} | | 0.0304 | [**S**$_\sigma$,**R**$_\sigma$] | {-0.037897[**R**$_\sigma$], 0.221850[**S**$_\sigma$], -0.003074[**R**$_\sigma$]} | | -0.0001 | [**R**$_\sigma$,**R**$_\sigma$] |
| $D_{5128}$ | {-0.113693[**R**$_\sigma$], -0.315810[**R**$_\sigma$], 0.003891[**S**$_\sigma$]} | | -0.0004 | [**R**$_\sigma$,**S**$_\sigma$] | {-0.804516[**R**$_\sigma$], -0.410188[**R**$_\sigma$], 0.050681[**S**$_\sigma$]} | | -0.0408 | [**R**$_\sigma$,**S**$_\sigma$] |

$\sum_{\mathbf{S\sigma,R\sigma}}\{\mathbb{C}_\sigma, \mathbb{F}_\sigma, \mathbb{A}_\sigma\}$  $\sum\mathbb{C}_{helicity}$  $\sum_{\mathbf{S\sigma,R\sigma}}\{\mathbb{C}_\sigma, \mathbb{F}_\sigma, \mathbb{A}_\sigma\}$  $\sum\mathbb{C}_{helicity}$

{2.2161 [**S**$_\sigma$], -0.0656 [**R**$_\sigma$], -0.0504 [**R**$_\sigma$]}  0.1239  [**S**$_\sigma$,**R**$_\sigma$]  {-2.2173 [**R**$_\sigma$], 0.0646 [**S**$_\sigma$], 0.0504 [**S**$_\sigma$]}  -0.1239  [**R**$_\sigma$,**S**$_\sigma$]

$\sum_{\mathbf{S\sigma}}\{\mathbb{C}_\sigma,\mathbb{A}_\sigma\}$  $\sum_{\mathbf{R\sigma}}\{\mathbb{C}_\sigma,\mathbb{A}_\sigma\}$  $\sum_{\mathbf{S\sigma}}\{\mathbb{C}_\sigma,\mathbb{A}_\sigma\}$  $\sum_{\mathbf{R\sigma}}\{\mathbb{C}_\sigma,\mathbb{A}_\sigma\}$

{2.4135 [**S**$_\sigma$], 0.1184 [**S**$_\sigma$]}  {-0.1973 [**R**$_\sigma$], -0.1688 [**R**$_\sigma$]}  {0.1973 [**S**$_\sigma$], 0.1688 [**S**$_\sigma$]}  {-2.4146 [**R**$_\sigma$], -0.1185 [**R**$_\sigma$]}

$\sum_{\mathbf{S\sigma}}\{\mathbb{C}_\sigma\}/|\sum_{\mathbf{R\sigma}}\{\mathbb{C}_\sigma\}|\}$ |$S_a$ = 12.2326

*Br-F-ethane*

| | | | | | | | | |
|---|---|---|---|---|---|---|---|---|
| $D_{3126}$ | {0.186670[**S**$_\sigma$], 0.033778[**S**$_\sigma$], 0.024561[**S**$_\sigma$]} | | 0.0046 | [**S**$_\sigma$,**S**$_\sigma$] | {-0.186563[**R**$_\sigma$], -0.033845[**R**$_\sigma$], -0.024561[**R**$_\sigma$]} | | -0.0046 | [**R**$_\sigma$,**R**$_\sigma$] |
| $D_{3127}$ | {0.463032[**S**$_\sigma$], 0.159524[**S**$_\sigma$], -0.043239[**R**$_\sigma$]} | | 0.0200 | [**S**$_\sigma$,**R**$_\sigma$] | {0.272965[**S**$_\sigma$], 0.712893[**S**$_\sigma$], -0.030877[**R**$_\sigma$]} | | 0.0084 | [**S**$_\sigma$,**R**$_\sigma$] |
| $D_{3128}$ | {-0.273034[**R**$_\sigma$], -0.712862[**R**$_\sigma$], 0.030841[**S**$_\sigma$]} | | -0.0084 | [**R**$_\sigma$,**S**$_\sigma$] | {-0.463089[**R**$_\sigma$], -0.160188[**R**$_\sigma$], 0.043228[**S**$_\sigma$]} | | -0.0200 | [**R**$_\sigma$,**S**$_\sigma$] |
| $D_{4126}$ | {0.154772[**S**$_\sigma$], 0.045147[**S**$_\sigma$], 0.027256[**S**$_\sigma$]} | | 0.0042 | [**S**$_\sigma$,**S**$_\sigma$] | {-0.168768[**R**$_\sigma$], -0.209019[**R**$_\sigma$], -0.030179[**R**$_\sigma$]} | | -0.0051 | [**R**$_\sigma$,**R**$_\sigma$] |
| $D_{4127}$ | {0.445947[**S**$_\sigma$], 0.187811[**S**$_\sigma$], -0.042255[**R**$_\sigma$]} | | 0.0188 | [**S**$_\sigma$,**R**$_\sigma$] | {0.328508[**S**$_\sigma$], 0.653119[**S**$_\sigma$], -0.025453[**R**$_\sigma$]} | | 0.0084 | [**S**$_\sigma$,**R**$_\sigma$] |
| $D_{4128}$ | {-0.223285[**R**$_\sigma$], -0.674241[**R**$_\sigma$], 0.028982[**S**$_\sigma$]} | | -0.0065 | [**R**$_\sigma$,**S**$_\sigma$] | {-0.279427[**R**$_\sigma$], 0.050013[**S**$_\sigma$], 0.048504[**S**$_\sigma$]} | | -0.0136 | [**R**$_\sigma$,**S**$_\sigma$] |
| $D_{5126}$ | {0.168652[**S**$_\sigma$], 0.209520[**S**$_\sigma$], 0.030182[**S**$_\sigma$]} | | 0.0051 | [**S**$_\sigma$,**S**$_\sigma$] | {-0.154738[**R**$_\sigma$], -0.045329[**R**$_\sigma$], -0.027212[**R**$_\sigma$]} | | -0.0042 | [**R**$_\sigma$,**R**$_\sigma$] |
| $D_{5127}$ | {0.279314[**S**$_\sigma$], -0.049058[**R**$_\sigma$], -0.048520[**R**$_\sigma$]} | | 0.0136 | [**S**$_\sigma$,**R**$_\sigma$] | {0.224020[**S**$_\sigma$], 0.674250[**S**$_\sigma$], -0.028989[**R**$_\sigma$]} | | 0.0065 | [**S**$_\sigma$,**R**$_\sigma$] |
| $D_{5128}$ | {-0.329263[**R**$_\sigma$], -0.653104[**R**$_\sigma$], 0.025360[**S**$_\sigma$]} | | -0.0083 | [**R**$_\sigma$,**S**$_\sigma$] | {-0.445767[**R**$_\sigma$], -0.187671[**R**$_\sigma$], 0.042154[**S**$_\sigma$]} | | -0.0188 | [**R**$_\sigma$,**S**$_\sigma$] |

$\sum_{\mathbf{S\sigma,R\sigma}}\{\mathbb{C}_\sigma, \mathbb{F}_\sigma, \mathbb{A}_\sigma\}$  $\sum\mathbb{C}_{helicity}$  $\sum_{\mathbf{S\sigma,R\sigma}}\{\mathbb{C}_\sigma, \mathbb{F}_\sigma, \mathbb{A}_\sigma\}$  $\sum\mathbb{C}_{helicity}$

{0.8728 [**S**$_\sigma$], -1.4535 [**R**$_\sigma$], 0.0332 [**S**$_\sigma$]}  0.0431  [**S**$_\sigma$,**S**$_\sigma$]  {-0.8729 [**R**$_\sigma$], 1.4542 [**S**$_\sigma$], -0.0334 [**R**$_\sigma$]}  -0.0430  [**R**$_\sigma$,**R**$_\sigma$]

$\sum_{\mathbf{S\sigma}}\{\mathbb{C}_\sigma,\mathbb{A}_\sigma\}$  $\sum_{\mathbf{R\sigma}}\{\mathbb{C}_\sigma,\mathbb{A}_\sigma\}$  $\sum_{\mathbf{S\sigma}}\{\mathbb{C}_\sigma,\mathbb{A}_\sigma\}$  $\sum_{\mathbf{R\sigma}}\{\mathbb{C}_\sigma,\mathbb{A}_\sigma\}$

{1.6984 [**S**$_\sigma$], 0.1672 [**S**$_\sigma$]}  {-0.8256 [**R**$_\sigma$], -0.1340 [**R**$_\sigma$]}  {0.8255 [**S**$_\sigma$], 0.1339 [**S**$_\sigma$]}  {-1.6984 [**R**$_\sigma$], -0.1673 [**R**$_\sigma$]}

$\sum_{\mathbf{S\sigma}}\{\mathbb{C}_\sigma\}/|\sum_{\mathbf{R\sigma}}\{\mathbb{C}_\sigma\}|\}$ |$S_a$ = 2.0572

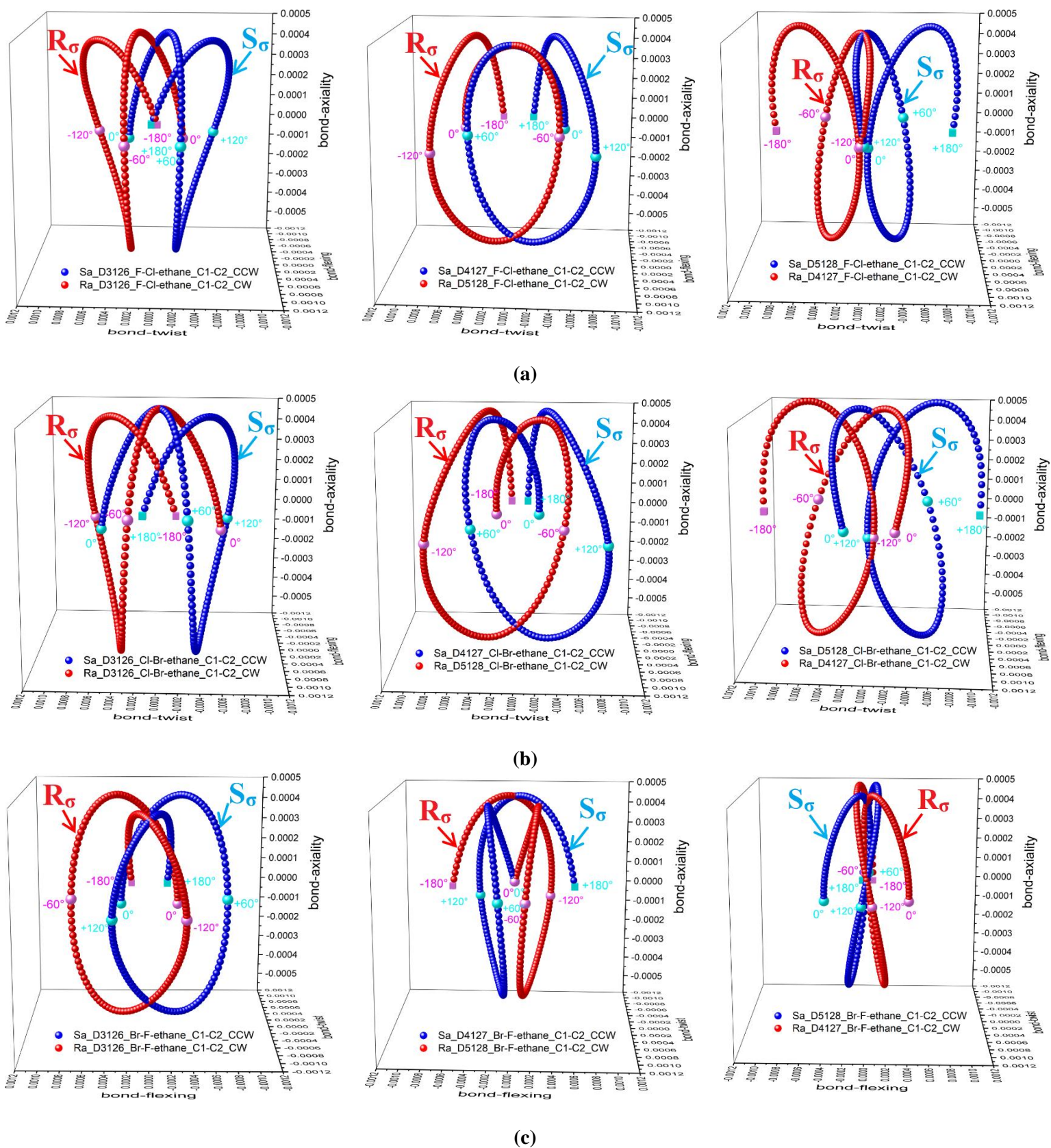

**Figure 1**. The doubly (F-Cl, Cl-Br and Br-F) substituted ethane C1-C2 *BCP* stress tensor trajectories $\mathbb{T}_\sigma(s)$ for the Cartesian CW and CCW torsions for $S_a$ and $R_a$ geometric stereoisomers of the **S**$_\sigma$ (blue) and **R**$_\sigma$ (red) $\mathbb{U}_\sigma$-space stereoisomers are presented in sub-figures **(a-c)** respectively, see **Table 2**.

**Conclusions**

In this NG-QTAIM chirality investigation, the singly and doubly halogen (F, Cl and Br) substituted ethane molecules were determined to be overall achiral and chiral in $\mathbb{U}_\sigma$-space respectively. The three singly substituted ethane molecular graphs were found to comprise $\mathbf{Q}_\sigma$ (the 'null' chirality assignment) along with $\mathbf{S}_\sigma$ and $\mathbf{R}_\sigma$ chirality $\mathbb{C}_\sigma$ assignments, but the total resultant contribution to the chirality-helicity function $\sum\mathbb{C}_{helicity} = 0$ indicated an overall lack of chiral character in $\mathbb{U}_\sigma$-space. The bond-axiality $\mathbb{A}_\sigma$ was found to respond much more strongly to the increase in atomic number of the singly substituted halogen than the chirality $\mathbb{C}_\sigma$. The F atom substitution resulted in significantly lower chirality $\mathbb{C}_\sigma$ contributions compared with the Cl and Br that suggest that the presence of F atom would increase the achiral characteristics of doubly substituted ethane.

Examination of the doubly halogen (F, Cl and Br) substituted ethane demonstrates agreement with the CIP naming schemes for all the $S_a$ and $R_a$ geometric stereoisomers for the dominant $\mathbb{U}_\sigma$-space $\mathbf{S}_\sigma$ and $\mathbf{R}_\sigma$ chirality stereoisomers. The additional feature provided by NG-QTAIM, compared with the CIP rules, is that we demonstrate there is a *mix* of $\mathbf{S}_\sigma$ and $\mathbf{R}_\sigma$ chirality for each of the $S_a$ and $R_a$ geometric stereoisomers. The presence of equal magnitude and opposite signs for the $\sum\mathbb{C}_{helicity}$ for the $S_a$ and $R_a$ geometric stereoisomers demonstrates consistency. In this investigation, on the basis of the lowest ratio for the $S_a$ geometric stereoisomer $\sum_{\mathbf{S}_\sigma}\{\mathbb{C}_\sigma/|\sum_{\mathbf{R}_\sigma}\{\mathbb{C}_\sigma\}|\}$, it is demonstrated that the F-Cl-ethane stereoisomers in $\mathbb{U}_\sigma$-space are the most achiral, closely followed by the Br-F-ethane stereoisomers, with the Cl-Br-ethane stereoisomers in $\mathbb{U}_\sigma$-space being the least achiral (most chiral). This indicates that the presence of the very light F atom, which is more similar to the hydrogen atom of the singly substituted ethane than Cl or Br, is responsible for the higher degree of achiral character present for the F-Cl-ethane and Br-F-ethane. This finding is consistent with the results from the singly (F) substituted ethane.

Future investigations of chiral and formally achiral molecules could be undertaken using $\mathbb{T}_\sigma(s)$ constructed from all possible torsion *BCP*s associated with any suspected chiral center. Further exploration of the newly discovered NG-QTAIM mixed chirality of stereoisomers in $\mathbb{U}_\sigma$-space could be undertaken by manipulating the degree of chiral/achiral character in $\mathbb{U}_\sigma$-space with, for instance, applied electric fields or laser irradiation. Reversing the $\mathbb{U}_\sigma$-space chirality with electric fields or laser irradiation fast enough to avoid disrupting atomic positions is in principle possible. Such a reversal of chirality in $\mathbb{U}_\sigma$-space would result in the $S_a$ and $R_a$ geometric stereoisomers predominately comprising $\mathbf{R}_\sigma$ and $\mathbf{S}_\sigma$ chirality assignments in $\mathbb{U}_\sigma$-space respectively. Reversing the $\mathbf{R}_\sigma$ and $\mathbf{S}_\sigma$ chirality may reasonably be expected to occur via a chiral to achiral transition in $\mathbb{U}_\sigma$-space.

**Funding Information**

The National Natural Science Foundation of China is gratefully acknowledged, project approval number: 21673071. The One Hundred Talents Foundation of Hunan Province is also gratefully acknowledged for the support of S.J. and S.R.K. H.F. and T.v.M. gratefully acknowledge computational support via the EaStCHEM Research Computing Facility.